# *Operando* XPS in Reactive Plasmas: The Importance of The Wall Reactions

J. Trey Diulus,[a]* Ashley R. Head,[b] Jorge Anibal Boscoboinik,[b] Andrei Kolmakov[a]*

[a.] Nanoscale Device Characterization Division, PML, NIST, Gaithersburg, MD, 20899, USA

[b.] Center for Functional Nanomaterials, Brookhaven National Laboratory, Upton, NY 11973

*john.diulus@nist.gov, andrei.kolmakov@nist.gov

**Abstract**

Advancements in differential pumping and electron optics over the past few decades have enabled x-ray photoelectron spectroscopy (XPS) measurements at (near-)ambient pressures, bridging the pressure gap for characterizing realistic sample chemistries. Recently, we demonstrated the capabilities of an ambient pressure XPS (APXPS) setup for *in-situ* plasma environment measurements, allowing plasma-surface interactions to be studied in operando rather than using the traditional before-and-after analysis approach. This new "plasma-XPS" technique facilitates the identification of reaction intermediates critical for understanding plasma-assisted surface processes relevant to semiconductor nanomanufacturing, such as physical vapor deposition, etching, atomic layer deposition, and many other plasma applications.

In this report, we apply the plasma-XPS approach to monitor real-time surface chemical changes on a model Ag(111) single crystal exposed to oxidizing and reducing plasmas. We correlate surface-sensitive data with concurrent gas-phase XPS measurements and residual gas mass-spectra analysis of species generated during plasma exposure, highlighting the significant role of plasma-induced chamber wall reactions. Ultimately, we demonstrate that plasma-XPS provides comprehensive insights into both surface and gas-phase chemistry, establishing it as a versatile and dynamic characterization tool with broad applications in microelectronics research. Finally, we outline potential enhancements and future metrology directions to advance plasma-XPS investigations further.

**INTRODUCTION**

Thanks to advancements in differential pumping and electron optics of electron energy analyzers over the past few decades[1-2], the pressure gap challenge in x-ray photoelectron spectroscopy (XPS) studies has been successfully resolved with the ambient pressure XPS (AP XPS, also known as NAP XPS) approach. The previous (ultra-)high vacuum operational limit was significantly extended to mbar region and even atmospheric pressure, allowing for *operando* studies of solid-gas, solid-liquid, and liquid-gas interfacial chemistry under realistic conditions (see comprehensive reviews [3-7] and references therein). Plasma-induced surface modification is a key technology in multiple applications spanning from semiconductor fabrication, aero-space industry, bio-medical treatments, environmental remediation, materials processing and fabrication, etc. While the importance of in-plasma surface analysis is well-recognized,[8-9] the application of surface-sensitive electron spectroscopies during plasma exposure has been largely delayed due to the aforementioned pressure gap and other experimental challenges. Some of the challenges can be and have been neatly resolved using spinning-wall[10] and through-the-membrane[11-12] methods. On the other hand, much of the semiconductor industry utilizes the plasmas in a pressure range spanning between $10^{-1}$ and $10^3$ Pa. This pressure range conveniently overlaps with the APXPS instrumental capabilities[6], making in-plasma XPS measurements feasible, provided XPS can still be collected under a plasma environment. Recently, we[13] and others[14] demonstrated this principal capability under (quasi-)remote plasma conditions using conventional APXPS equipment, and the plasma-XPS method is currently gaining momentum.[15]

Utilizing operando plasma-XPS for surface and gas phase analysis along with mass spectrometry, as demonstrated in this report, provides an ideal route to identifying reaction intermediates or meta-stable surface species that are difficult to measure using the standard before-after method. Using a model Ag(111) single crystal, we comparatively assessed Ag surface reduction and oxidation at RT upon exposure to a low-power (15 W) quasi-remote plasma in oxidizing and reducing environments. Complementary residual gas analysis (RGA) provides a chemical fingerprint of the molecular products during the plasma-induced reaction and is further verified with gas-phase APXPS. Ultimately, we demonstrate the ability to track dynamic changes in the surface chemistry related to global plasma-induced processes that are potentially valuable for semiconductor process development and control.

**METHODS**

Plasma-XPS experiments were performed at the Center for Functional Materials at Brookhaven National Laboratory equipped with a lab-based APXPS system. The APXPS instrument (Figure 1a) utilizes a reaction chamber with a base pressure of $<5\times10^{-7}$ Pa and the ability to backfill the entire chamber with hydrogen and oxygen gas during data collection.[16] A single crystal Ag(111) was mounted to a sample holder via spot welds and loaded into the APXPS chamber. The sample was not sputtered or annealed in order to begin with a surface that has adventitious species present to determine the "cleaning" effect of the initial $O_2$ plasma exposure. This Ag crystal is typically used to monitor the photocurrent of the monochromatized X-ray source, providing a metric for beam focusing and source anode lifetime. The photocurrent is measured by removing the ground connector from the manipulator and then connecting it to an electrometer. Generally, the photocurrent for a well-focused X-ray source is within the $100 \pm 10$ pA range.

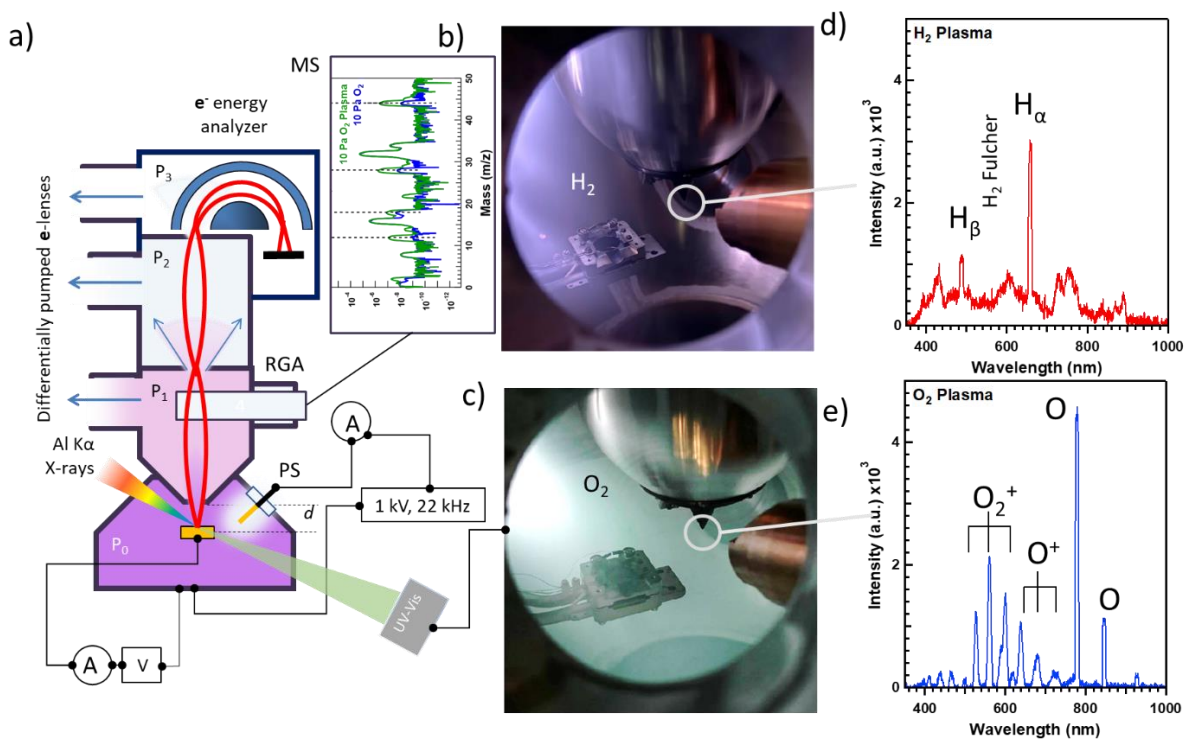

**Figure 1.** a) Experimental setup of the plasma-XPS with three differential pumping stages. The sample is connected to an electrometer and also can be biased. PS, RGA, and MS correspond to the plasma source, residual gas analyzer, and mass spectrum, respectively. Images (b and c) as well as corresponding optical emission spectra (d and e) of hydrogen and oxygen plasmas ignited in the analysis chamber with the sample stage retracted (out of the XPS focal position). The circles in the b) and c) panels show the approximate position of the spectra collection point.

**Plasma parameters**

Plasma was generated in the APXPS by backfilling the analysis chamber to 10 Pa of either $H_2$ or $O_2$ separately and applying 1 keV peak-to-peak AC voltage (22 kHz) to an in-chamber copper electrode mounted on a high-voltage feedthrough (Figure 1a). Figure 1 displays the images of the plasma glows (Figures 1b and c) and the corresponding optical emission spectra of both plasmas (Figures 1d and e). The high voltage power supply was consistently set to a power output corresponding to approximately 15 W. The exact power flux of the plasma will be dependent on the gas used to ignite the plasma, but for the scope of this paper, we refer to setting the power output of the current setup.

During XPS spectra acquisition, the sample was located ca 0.6 mm in front of the grounded analyzer cone and 10 cm away from the driving copper electrode (Figure 1b and d). Thus, the sample is under (quasi-)remote plasma conditions where the concentration of ions and electrons is low, and surface redox reactions are controlled mainly by neutral radical species.[13] To get a sense of the plasma conditions, we use optical emission spectroscopy, shown both for the $H_2$ (Figure 1b) and $O_2$ (Figure 1c) plasmas. The optical spectra were collected from the plasma volume near the sample using a long focal-length lens. Therefore, the spectrum measured should represent the particle environment seen by the sample. An initial "dark" scan was collected by completely covering the lens to subtract it from the plasma spectra.

For both gases, the spectra were dominated by atomic lines, which is typical for inductively coupled plasmas in this pressure range.[17-18] In the $O_2$ plasma emission spectrum, a set of peaks representing $O_2^+$ (526 nm, 560 nm, and 600 nm) and $O^+$ (638 nm, 680 nm, and 723 nm), are seen along with two O radical peaks, 778 nm and 845 nm. The measurable intensity below 450 nm could be due to CO and $CO_2^+$ emission bands[18] (see also discussion below). The $H_2$ plasma emission spectrum (Figure 1d) shows the major atomic recombination peaks $H_\alpha$ and $H_\beta$, in addition to the appreciably lower intensity of the broad molecular ($H_2$ Fulcher) band. This is confirmed by a hydrogen plasma afterglow that exhibits a violet hue, seen in Figure 1b, due to the mixing of H-alpha ($H_\alpha$) and H-beta ($H_\beta$) Balmer series visible spectral lines of the hydrogen atom at 656 nm and 486 nm, respectively. For the $H_2$ plasma, the plume is more concentrated near the driving Cu electrode, and emission is weak near the sample, while the $O_2$ plasma glow seemingly propagates over the entire chamber. When collecting the spectra, the $H_2$ integration time was set to 10 seconds, while the $O_2$ integration time was lowered to 1 second to avoid saturation of the detector. When plotting the optical spectra in Figure 1, the $O_2$ data are normalized by a factor of 10, and both are plotted to the same y-axis range to display a relative band's intensities. Overall, for the same pressure of 10 Pa and driver power, the $H_2$ plasma seems to be less intense.

**APXPS Parameters**

The reaction chamber is separated from the multi-stage differentially pumped electrostatic focusing lens system of an electron spectrometer by a 300 μm diameter cone aperture to enable XPS data collection at pressures up to ca. 200 Pa. The sample position was adjusted to the focal point of the X-rays (600 μm below the aperture) by optimizing the intensity of a photoemission peak. A monochromatized Al Kα X-ray source (hν = 1487 eV), focused to a ca. 300 μm diameter spot size and fixed at 55° from the sample normal, was used for acquiring XPS spectra. Survey spectra were collected using a pass energy of 50 eV with a dwell time of 100 ms and a step size of 1 eV. High-resolution spectra were collected with a 20 eV pass energy, 250 ms dwell time, 50 meV step size, and sufficient sweeps for a better signal-to-noise ratio, with a doubled number of sweeps for data collected at elevated pressure. No effect of plasma ignition on photoelectron signal noise was noticed. A standard set of scans would consist of a survey, Ag 3d, O 1s, C 1s, and valence band, which would total roughly 2000 seconds of exposure time. An additional 830 seconds would be added if the collection of gas phase spectra was warranted. Thus, the sample was exposed to plasma for approximately 50 min, accounting also for initial plasma ignition and scan parameter setup.

Gas phase XPS were collected with the sample retracted downward in the z position, thus away from the focus of the x-ray source and analyzer, but still underneath the cone. In this way, the gas phase probed is directly above the sample and presumably has a fraction of the molecules that interact or scatter with/at the sample surface. Still, we are probing a gaseous mixture that will contain background species that are formed due to reactions with chamber walls and this signal is difficult to differentiate from the resulting chemistry that takes place on the sample surface. The gas phase signal-to-noise ratio is substantially lower than data collected with a solid sample due to the low concentrations of molecules at the plasma pressures. Gas phase XPS were also collected with no sample present, which is related exclusively to the plasma gas itself and products of the chamber walls reactions. However, no systematic studies have been performed yet to isolate gas phase contribution related to plasma-sample reactions. Therefore, the gas phase XPS is only used here to identify the commutative molecular content in front of the XPS analyzer nozzle.

**RGA Parameters**

The first APXPS differential pumping stage is equipped with a quadrupole mass spectrometer for residual gas analysis of the reaction chamber environment during plasma exposure (Figure 1a). The electron multiplier was used to collect residual gas analysis (RGA) mass spectra (MS) when the pressure in the first stage fell below $10^{-4}$ Pa. A survey MS scan was collected for a mass-to-charge ratio (m/z) of 1 Da/e to 50 Da/e with a sweep time of ca. 33.5 seconds during APXPS data collection. The data were collected as a continuous scan during multiple conditions: (1) initial APXPS scans under ultra-high vacuum (UHV) conditions, (2) as the chamber was backfilled to 10 Pa of $O_2$, (3) after the chamber was purged back to UHV, (4) as the chamber was backfilled to 10 Pa of $H_2$, and 5) after the chamber was purged again to UHV. To differentiate the on-sample reactions from the side (on-walls) reactions, RGA was collected with the same five step procedure but with no sample present to simulate the same environment where the contribution from the background can be subtracted to highlight the sample contribution.

**RESULTS and DISCUSSION**

XPS data and RGA spectra were collected simultaneously at each pressure condition and plasma environment. Once the plasma was ignited, optical spectra were collected while XPS and RGA scans were running. When the XPS scan finished, the sample was moved away from the analyzer focal point to collect exclusively gas phase XPS spectra, and then a new chamber pressure condition was set. Here, we focus on the oxidation and reduction of the Ag surface during plasma exposure. Note that the loaded Ag sample was exposed to the atmospheric ambient, and its surface presumably is pre-covered with the carbonaceous species prior to plasma "cleaning" reactions. In short, the results demonstrate the initial "combustion" of impurity carbon by oxygen plasma is followed by oxidation of the surface, which can then be reduced back to the $Ag^0$ state by exposure to hydrogen plasma. This plasma-induced spectra progression generally follows the prior XPS Ag oxidation/reduction results[19-25] and also recent APXPS observations[14] and is highlighted in the measurements series (from the bottom row to the top row sequence) shown in Figure 2.

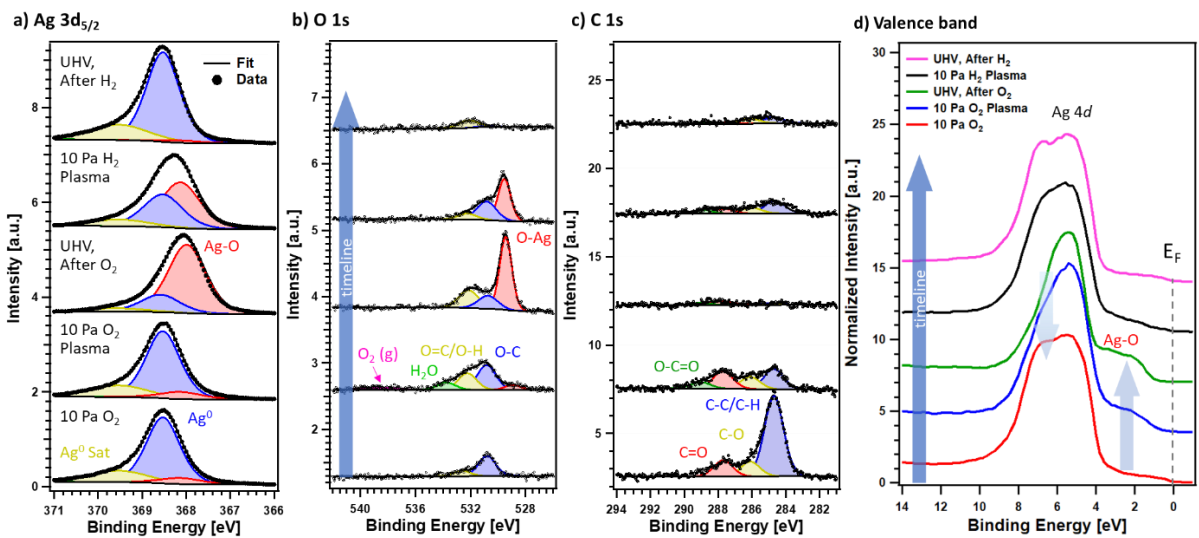

**Figure 2.** XPS of the oxidation and reduction of Ag sequence from the bottom to the top rows: bottom: exposed to 10 Pa $O_2$; second: during 10 Pa $O_2$ plasma treatment; after quenching $O_2$ plasma but still with overpressure of $O_2$ (third); during 10 Pa $H_2$ plasma treatment (fourth row), and after quenching $H_2$ plasma still with overpressure of $H_2$ (top) for the Ag 3d (a), C 1s (b), and

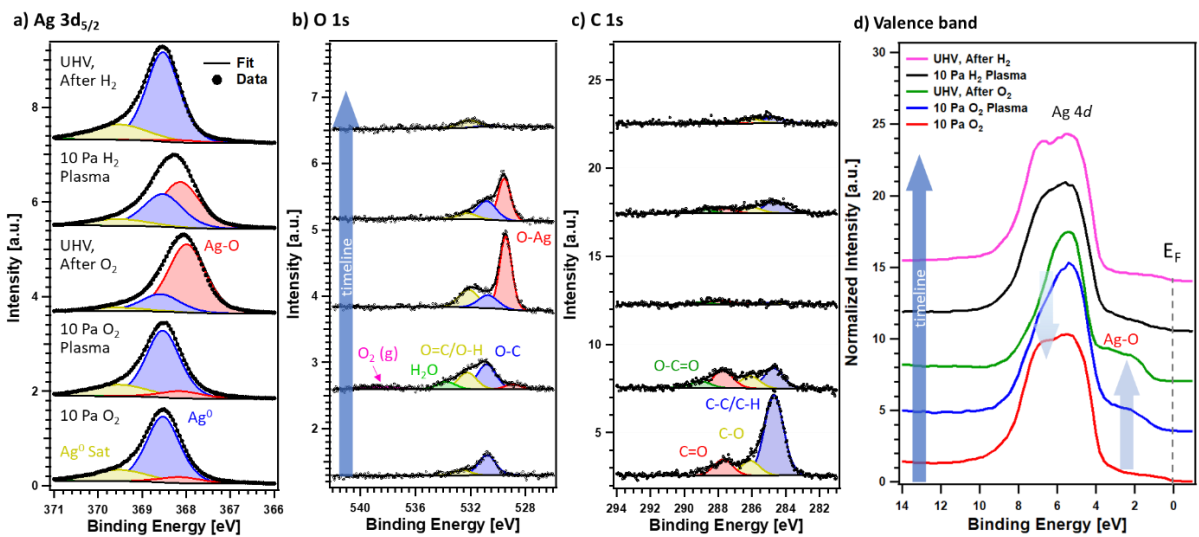

O 1s (c) core levels. Additionally, the corresponding valence band spectra evolution is shown in (d).

**Oxidation of Ag**

Prior to plasma-induced oxidation, the loaded Ag sample had a significant amount of carbon on the surface (see bottom panels in Figure 2). Under exposure to 10 Pa $O_2$ (left bottom panel in Figure 2), the Ag 3d showed no distinct evidence of oxidized Ag, suggesting the peak recorded in the O 1s at 530.8 eV mainly corresponds to oxidized carbon impurities.[19, 21] This is most likely adsorbed alcohol or ketone species that are also seen as 285 eV and 287 eV bands in the C 1s. The gas phase C 1s shown in Figure 3b has no signal above the background, while the O 1s gas phase (Fig. 3a) shows a standard doublet corresponding to the paramagnetic nature of molecular $O_2$.[26]

As the plasma is ignited at 10 Pa (second row from the bottom in Figure 2) the Ag 3d shape stays mostly the same initially, with no evidence of oxidation. A substantial drop in the aliphatic carbon peak at 284.8 eV is seen, while a clear increase in the oxidized carbon is seen. The gas phase O 1s also shows an additional peak at 532 eV, which can correspond to $CO_2$, but also may be due to molecules containing OH species as a result of the $O_2$ plasma reacting with residual water on chamber walls. The gas phase C 1s shows (Fig.3 b) a clear appearance of CO [27] and $CO_2$ [28] peaks above the background. The gas phase O 1s also shows a broadening of the doublet; however, this is most likely not a contribution of CO, $CO_2$, or $H_2O$, as all three of these should be at lower energies than the initial doublet peak of $O_2$. Instead, the variation of the local plasma potential seems to be responsible for the broadening effect and appreciable (~1 eV) shifts to higher binding energies of the molecular $O_2$ spectrum. The plasma potential has a positive value in front of the sample with respect to the ground, and it depends on multiple parameters, such as geometry and the areal ratio between the plasma driving electrode and chamber walls (including the analyzer cone), pressure, gas type, driving voltage, etc. This also explains why the spectra intensity drops upon plasma ignition and all spectra generally shift to higher binding energy. The origin and values of the binding energy shifts in the plasma environment will be described in detail in the forthcoming report.

Following the prolonged exposure to plasma, the sample returns to an $O_2$ overpressure of 10 Pa, corresponding to the third row (from the bottom) in Figure 2. Now, after ca. 40 min of oxidizing plasma exposure, the Ag 3d shows a broadened peak with a slight shift in the total peak to lower BE. A clear sharp peak in the O 1s at 529.5 eV corroborates with a metal oxide peak, suggesting the formation of a stable surface oxide. The broadening of the gas phase O 1s doublet has also been reduced as the plasma is quenched, although there appears to be a discharge effect that still provides slightly more broadening to the doublet than prior to igniting the plasma. Finally, there is essentially no C left after the extended $O_2$ plasma exposure, suggesting the removal of surface adsorbates.

The evolution of the valence band (VB) spectra follows the trends above (Figure 2 right panel). The initial spectrum (red curve) collected under 10 Pa of $O_2$ shows a typical Ag (111) VB shape defined by photoemission from Ag 4d (major doublet band) and 5 s states (shallow band starting at $E_F$). Oxygen plasma treatment (blue and green curves) results in the appearance of the broad unresolved emission band between 1 eV and 4 eV below $E_F$, which is commonly assigned to the formation of strongly Ag-bound oxygen states.[19-20, 22, 29] In addition, the higher energy side of the 4 d band loses its intensity similar to prior results on $Ag_2O$ formation and reduction on Ag films.[30]

**Reduction of Ag**

To reduce the surface of Ag following plasma oxidation, the chamber was pumped to UHV and then filled to 10 Pa but now with $H_2$ gas. No obvious changes in Ag 3d, O 1s, and C 1s can be seen with just exposure to molecular $H_2$ (not shown here). After that, hydrogen plasma was ignited, and spectra were collected again, corresponding to the fourth XPS spectra row from the bottom in Figure 2. Immediately noticeable is a slight increase in C 1s, which must come from the $H_2$ plasma scrubbing chamber walls and volatilizing carbon species that are then re-deposited onto the surface of Ag. The O 1s also shows a slight decrease in the oxidized metal peak at 529.5 eV, which corresponds to a slight shift back to 368.1 eV in the Ag 3d.

Unfortunately, the comparatively low concentration of hydrogen plasma-induced volatile reaction products resulted in no measurable signal in gas phase XPS at either C 1s or O 1s ranges (Figure 3). This is not surprising since prior comparative studies of low-pressure plasmas reveal an order of magnitude reduction of densities of hydrogen radicals compared to oxygen ones under the same power and pressure conditions. [31-32]

For the Ag sample, the overall exposure time to hydrogen plasma was ca. 40 min, and then spectra were collected with still an overpressure of 10 Pa $H_2$, corresponding to the top spectra in

Figure 2. Following $H_2$ plasma treatment, the Ag 3d has returned to 368.3 eV and no evidence of a metal oxide is seen in the O 1s. The C 1s show approximately the same amount of C at the initial $H_2$ plasma exposure, suggesting that hydrogen plasma is removing carbon-containing species from the Ag surface not as effectively as oxygen plasma, but is able to reduce the metal surface.

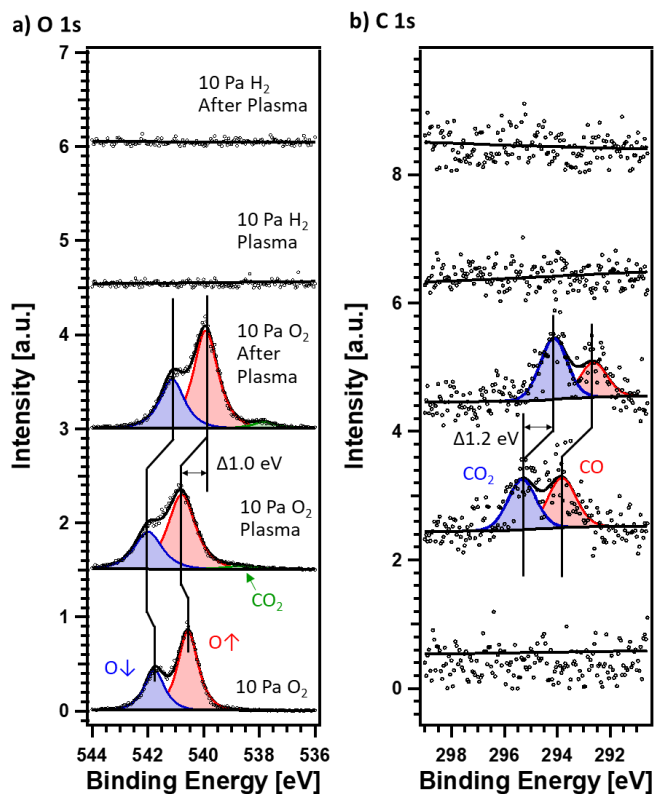

**Figure 3.** Gas phase XPS of both the O 1s (a) and C 1s (b) core levels for initial backfilling of molecular $O_2$ in the analysis chamber (bottom row), followed by plasma ignition (second raw from the bottom) and quenching with the same chamber environment (third row). Then the chamber is purged and backfilled with the same pressure of molecular $H_2$ gas, followed by $H_2$ plasma ignition (fourth row from the bottom). The binding energy shift upon plasma ignition is due to plasma potential build-up near the analyzer cone orifice.

The valence band spectra (Figure 2 right panel) support the reduction of the $Ag_xO$ layer upon hydrogen plasma treatment (black and pink curves), where the disappearance of oxygen-induced states centered around 3 eV and recovery of the clean Ag 4d band can be observed.

**Residual Gas Analysis**

In order to support XPS data on the reactions taking place during plasma exposure, we utilize the RGA located behind the analyzer cone. In this case, we assume, that due to the proximity of the cone to the sample for XPS analysis (600 μm) and the cone orifice (300 μm), the large fraction of the molecules detected by the RGA are molecules scattered by the sample surface together with the volatile reaction products originated from the sample surface. Again, the exact partitioning of the observed data between the sample surface and the chamber wall remains a subject of future studies. In our prior work [13], we demonstrated a mass-spectra analysis where we collected the evolutions of only seven specific masses with respect to time to highlight the significance of plasma-induced volatile species from the walls reactions. However, this method limits the full understanding of all potential products to the ones we expect. Therefore, here we utilized successive survey mass-spectroscopy scans from 1 amu to 100 amu to better showcase reaction products.

Figure 4 shows RGA data plotted on a log scale and then offset to show a progressive change. Initially, we collect a UHV baseline, shown in red, followed by an increase in $O_2$ overpressure to 10 Pa, shown in blue. The initial background shows a small amount of water background, which is expected due to the analysis chamber being unbaked. The introduction of $O_2$ shows a clear increase in m/z 32 ($O_2^+$), 16 ($O^+$) and 8 ($O^{2+}$), but also there is a small amount of CO and $CO_2$ from m/z 28 and 42 that may be from outgassing from chamber walls[33], but still 3-4 orders of magnitude lower than the $O_2$ signal. When the $O_2$ plasma is ignited, as shown by the green survey, a clear increase in the CO and $CO_2$ signals is seen, in addition to a substantial increase in m/z 12, which is a cracking signature of both CO and $CO_2$. This observation agrees well with the APXPS results that show the presence of these species in the gas phase spectra (Figure 3b), along with the uptake of these species adsorbed on the sample surface (Figure 2).

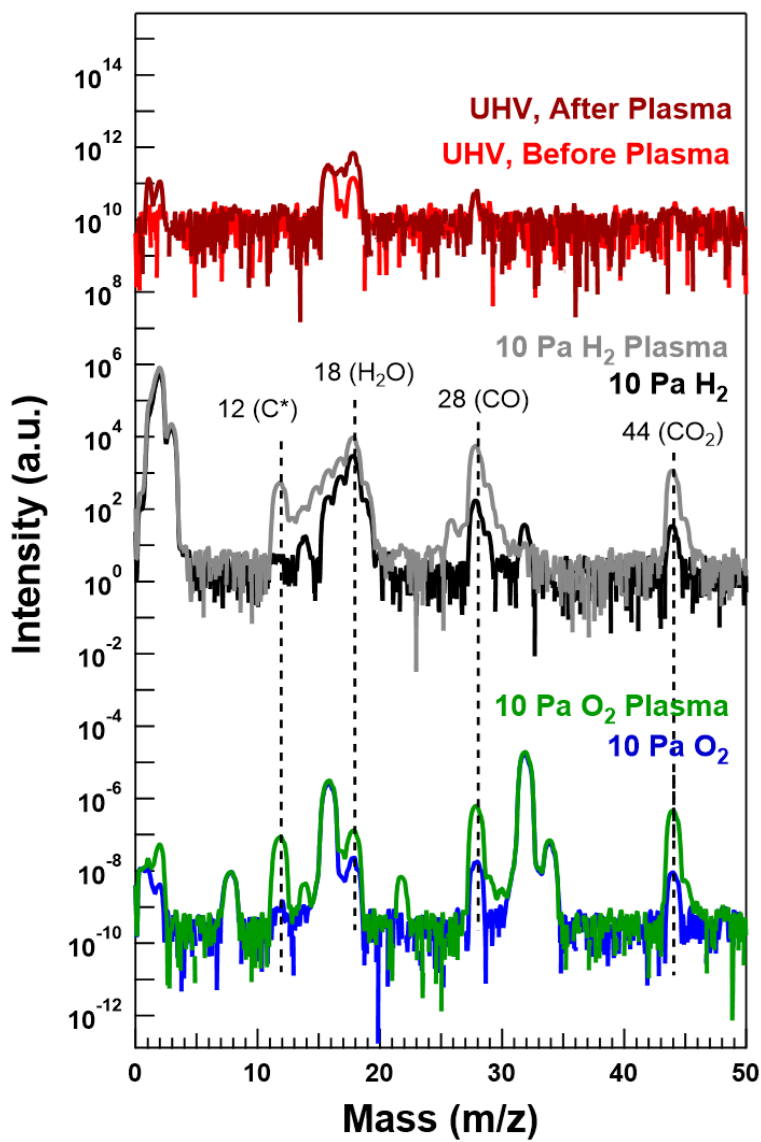

**Figure 4.** RGA surveys plotted on a $\log_{10}$ y-axis scale and offset to highlight the change with respect to chamber condition, starting with UHV before plasma (red, bottom), 10 Pa of $O_2$ (blue), $O_2$ plasma with 10 Pa (green), 10 Pa of $H_2$ (black), $H_2$ plasma with 10 Pa (grey), and after returning to UHV (red, top).

After the $O_2$ plasma, the chamber is pumped to UHV, and 10 Pa of $H_2$ is introduced, corresponding to the black survey. A background of water is still seen, although about 3 orders of magnitude lower than the $H_2$ signal. Even lower is a background of $O_2$, CO, and $CO_2$, again likely due to outgassing upon backfilling the chamber[33] and previous plasma exposure. As the hydrogen plasma is ignited, as shown by the grey survey, again, the CO and $CO_2$ increase similarly to the $O_2$ plasma case. This was interesting because the APXPS showed no evidence of these species in the gas phase, although clearly there was C seen on the Ag surface. However, the overall intensity of both CO and $CO_2$ was lower during $H_2$ plasma compared to the $O_2$ plasma. There was also a slight increase in $H_2O$ from m/z 18 that provides another reactive route to the removal of oxide from the surface.

Overall, when the chamber returns to UHV, shown by red again at the top of Figure 4, a clear increase in overall $H_2O$ background is seen, which strongly suggests the production of water took place during $H_2$ plasma exposure. Although we were unable to detect the volatile species with gas phase XPS during $H_2$ plasma, we still see a clear surface reduction corroborating the plasma interactions and stimulations of the reactions at the Ag surface. One must also consider the pumping efficiency and sticking probability of each gas, where water takes a longer time to pump compared to gasses like CO and $CO_2$. In this case, we likely overestimate the concentration of the carbonaceous species in the RGA relative to the produced water, which sticks to the walls more readily upon changing the pressure equilibrium.

**CONCLUSIONS AND OUTLOOK**

With this work, we continue introducing Plasma XPS metrology and demonstrate its capabilities. The results presented here demonstrate the powerful capabilities of ambient pressure x-ray photoelectron spectroscopy (APXPS) for studying surface chemistry in reactive plasma environments, overcoming the limitations of traditional before-and-after analytical techniques. By employing the plasma-XPS approach, we successfully monitored real-time oxidation and reduction processes on a model Ag(111) surface. The findings highlight the effectiveness of plasma-induced chemistry, including the efficient removal of adventitious carbon during oxygen plasma exposure and subsequent reduction of surface oxides using hydrogen plasma.

Our study also revealed the important role of plasma-induced chamber wall reactions, demonstrating that contributions from these secondary processes can complicate direct interpretations of surface chemistry and should be considered during real-world applications. Simultaneous residual gas analysis (RGA) provided complementary insights, confirming the formation of plasma-induced gas-phase reaction products such as CO and $CO_2$ originating from the chamber walls. These results underscore the importance of coupling surface and gas-phase analysis to fully understand complex plasma-surface interactions.

Moreover, we have demonstrated, for the first time, that gas phase XPS spectra (intensity, BE shifts, and peak broadness) recorded in a plasma environment are dependent on plasma parameters. Therefore, the chemical composition of the plasma itself and plasma potential can be accessed, providing a complementing powerful plasma diagnostics method.

Future investigations should prioritize isolating and quantifying wall effects to refine surface data interpretation further. Enhanced wall conditioning/cleaning protocols, combined with systematic gas-phase XPS studies under controlled plasma conditions, will improve the robustness of plasma-surface interaction models. Looking forward, several avenues for advancing plasma-XPS research merit exploration. First, integrating traditional plasma diagnostic tools and surface-sensitive optical spectroscopy methods with APXPS will allow for a more precise characterization of plasma-surface interactions and modeling. Second, extending the range of studied materials beyond simple metal surfaces will broaden the applicability of the technique to complex systems relevant to advanced semiconductor processes. Third, dynamic plasma diagnostics combined with modeling and real-time XPS measurements both in the gas phase and at the surface could enable kinetic studies of rapid surface transformations, offering a pathway to optimize industrial plasma applications, including atomic layer deposition and plasma etching.

Finally, advancements in APXPS instrumentation, such as improved differential pumping designs, more efficient photoelectron extracting optics, and enhanced detector sensitivity/acquisition rate, will facilitate the exploration of plasma-surface interactions at even higher pressures and more industry-relevant reactive environments. Collectively, these improvements have the potential to establish plasma-XPS as a valueable technique for operando studies in plasma-assisted manufacturing and surface chemistry research.

## ACKNOWLEDGMENTS AND DISCLAIMERS

The authors are grateful to Dr. Thomas P. Moffat and Dr. Sujitra J. Pookpanratana (both at NIST) for their careful reading of the manuscript and critical suggestions. This work was performed with funding from the CHIPS Metrology Program, part of CHIPS for America, National Institute of Standards and Technology, U.S. Department of Commerce. CHIPS for America has financially supported this work through the "Multiscale Modeling and Validation of Semiconductor Materials and Devices project". Any mention of commercial products in this article is for information only; it does not imply recommendation or endorsement by NIST. This research used the Proximal Probes Facility of the Center for Functional Nanomaterials (CFN), which is a U.S. Department of Energy Office of Science User Facility, at Brookhaven National Laboratory under Contract No. DE-SC0012704.

**AUTHOR DECLARATIONS**

The authors have no conflicts to disclose.

**AUTHOR CONTRIBUTIONS**

J. Trey Diulus: Data curation (lead): Formal analysis (lead); Investigation (lead); Writing – review & editing (lead). Ashley Head and Jorge Anibal Boscoboinik: formal analysis (supporting); Investigation (supporting); Writing – review & editing (supporting). Andrei Kolmakov: Conceptualization (lead); Formal analysis (equal); Investigation (equal); Supervision (lead); Writing –review & editing (lead).

**DATA AVAILABILITY**

The data that support the findings of this study are available from the corresponding author upon reasonable request.